\documentclass{ADAPTCONF2025}

\usepackage{placeins}
\usepackage[T1]{fontenc}
\usepackage{listings}
\usepackage{balance}

\interspeechcameraready

\title{Formalising Software Requirements with Large Language Models}
\name{Arshad Beg$^1$, Diarmuid O'Donoghue$^1$, Rosemary Monahan$^1$
}

\address{
  $^1$Department of Computer Science, Maynooth University, Ireland}
\email{arshad.beg@adaptcentre.ie, diarmuid.odonoghue@mu.ie, rosemary.monahan@adaptcentre.ie}

\begin{document}

\maketitle
\begin{abstract}
This paper is a brief introduction to our recently initiated project named VERIFAI: Traceability and verification of natural language requirements. The project addresses the challenges in the traceability and verification of formal specifications through providing support for the automatic generation of the formal specifications and the traceability of the requirements from the initial software design stage through the systems implementation and verification. Approaches explored in this project include Natural Language Processing, use of ontologies to describe the software system domain, reuse of existing software artefacts from similar systems (i.e. through similarity based reuse) and large language models to identify and declare the specifications as well as use of artificial intelligence to guide the process.
\end{abstract}

\section{Introduction}

Overlooking the formal requirements stage in software development often leads to ambiguous requirements. Requirements which are not expressed in a formal mathematical notation cannot have their correctness guaranteed through formal verification techniques which are required to maintain standards in safety critical software. These requirements are written informally and their correctness is not ensured. To gather them formally using some formal notation needs training of the software developers in the domain of requirements engineering and especially formal methods. It increases software development cycle time by a factor of 30\% \cite{huisman2024formalmethodsacademiaindustrial}. There is a need to bridge the gap between the work of the formal methods research community and the software industry, where the practices are rarely followed due to the fast pace of industry environment i.e. meeting deadlines and launching new software products.  

Formalising software requirements ensures clarity, correctness and verifiability. Formalisation involves development of formal languages, logic and verification techniques i.e. theorem proving and model-checking. Like all other fields, the development of Large Language Models (LLMs) opened a world of opportunities for the task of formalisation of software requirements. We can exploit the power of these LLMs to generate formal requirements and specifications. 

In this short paper, we shall represent an example of generating formal requirements and specifications from the existing literature in section \ref{sec:example}. Furthermore, we present the key papers and their contribution in this domain in section \ref{sec:relatedwork}. 

\section{Example of Generating Assertions}
\label{sec:example}

Dafny is a verification-aware programming language that has native support for expressing formal specifications and is equipped with a static program verifier to automatically verify implementations against specifications.  \cite{mugnier2024laurelgeneratingdafnyassertions} presented a framework named Laurel to generate Dafny assertions using LLMs. Mugnier et al. \cite{mugnier2024laurelgeneratingdafnyassertions} designed two domain-specific prompting techniques. The first one locates the position in code where assertion, which provide part of the formal specifications, is missing. This is done through analysis of the verifier's error message. At the particular location with missing assertion, a placeholder is inserted. A second technique involves provision of example assertions from a codebase. Laurel was able to generate over 50\% of the required helper assertions, making it a viable approach to deploy, while automating program verification process. 

\subsection{Example of Generating Dafny Assertions}

We illustrate an example of a Dafny lemma as reported in \cite{mugnier2024laurelgeneratingdafnyassertions}, where a helper assertion is needed in the verification process. The lemma ensures that the integer and fractional parts are correctly extracted while parsing a decimal string.

\subsection{Dafny Lemma Definition}

The following Dafny lemma, \texttt{ParseDigitsAndDot}, specifies the expected behaviour of the function \texttt{ParseDecStr}. It operates on an input string composed of digits and a decimal separator.


\smallskip  
\noindent\textbf{Lemma Definition:}  

\texttt{lemma ParseDigitsAndDot(s1: string, s2: string)}  

\texttt{\quad \textbf{requires} $\forall$ i | 0 $\leq$ i $<$ |s1| :: '0' $\leq$ s1[i] $\leq$ '9'}  

\texttt{\quad \textbf{ensures} ParseDecStr(s1+"."+s2).value.1 == "."+s2}  

\texttt{\quad \{}  

\texttt{\quad \quad if |s1| == 1 \{}  

\texttt{\quad \quad \quad \textbf{assert} ParseDecStr("."+s2).None?;}  

\texttt{\quad \quad \} else \{}  

\texttt{\quad \quad \quad ParseDigitsAndDot(s1[1..],s2);}  

\texttt{\quad \quad \quad \textbf{assert} s1 + "." + s2 == [s1[0]] + (s1[1..] + "." + s2);}  

\texttt{\quad \quad \}}  

\texttt{\quad \}}  

\subsubsection{Explanation of the Lemma}

- \textbf{Precondition} (\textbf{requires} clause): The function assumes that every character in \texttt{s1} is a digit ('0' to '9').  \\
- \textbf{Postcondition} (\textbf{ensures} clause): The function guarantees that when \texttt{ParseDecStr} is applied to the concatenation of \texttt{s1} and \texttt{s2}, the first part of the result remains \texttt{s1}, and the second part contains \texttt{"." + s2}.  \\
- \textbf{Base Case}: If \texttt{s1} contains a single character, the function asserts that parsing \texttt{"."+s2} leads to an empty integer part.  \\
- \textbf{Recursive Case}: The function calls itself with the tail of \texttt{s1} to process it recursively. However, to help the verifier understand the transformation, an assertion is added:  
  \begin{quote}
  \texttt{\textbf{assert} s1 + "." + s2 == [s1[0]] + (s1[1..] + "." + s2);}
  \end{quote}
  This assertion explicitly states how the string is decomposed and aids the SMT solver in proving correctness.

\subsubsection{Role of the Helper Assertion}

Without the assertion, the Dafny verifier struggles to establish the correctness of the postcondition due to the complexity of reasoning about string concatenation. The assertion serves as an intermediate step, breaking down the transformation into a form that is easier for the solver to handle.

This example demonstrates the importance of inserting helper assertions in Dafny proofs. Tools like Laurel \cite{mugnier2024laurelgeneratingdafnyassertions} aim to automate this process by leveraging Large Language Models to suggest relevant assertions.

\section{Related work}
\label{sec:relatedwork}

The authors of \cite{cosler2023nl2specinteractivelytranslatingunstructured} describe nl2spec, a framework that leverages LLMs to generate formal specifications from natural language. An open-source implementation with a web based interface is provided to support users in iteratively refining translations, making formalisation easier. 

AssertLLM tool \cite{10691792} exploits three customised LLMs to generate assertions for the hardware verification of design specifications. It is done in three phases, first understanding specifications, mapping signal definitions and generating assertions. The results show that AssertLLM produced 89\% correct assertions with accurate syntax and function. SpecLLM \cite{li2024specllmexploringgenerationreview} explores the space of generating and reviewing VLSI design specifications with LLMs. The cumbersome task of chip architects can be improved by exploiting the power of LLMs for synthesising natural language specifications involved in chip designing. Here, the utility of LLMs is explored with the two stages i.e. (1) \textbf{generation} of architecture specifications from scratch and from register transfer logic (RTL) code; and (2) \textbf{reviewing} these generated specifications. 

The Java Modelling Language (JML) is used for the formal specification and verification of Java programs. In \cite{10207159}, symbolic NLP and ChatGPT performance is compared in their generation of correct JML from given natural language specifications. The work \cite{quan2024verificationrefinementnaturallanguage} provides verification and refinement of natural language explanations by making LLMs and theorem provers work together. A neuro-symbolic framework i.e. Explanation-Refiner is represented. LLMs and theorem provers are integrated together to formalise explanatory sentences. The theorem prover then provides the guarantee of validated sentence explanations. Theorem provers also provide feedback for further improvements in the NLI (Natural Language Inference) model. Error correction mechanisms can also be deployed by using the tool Explanation-Refiner. 

Granberry et al. \cite{Granberry2025a} explored how combining large language models (LLMs) with symbolic analysis can help generate specifications for C programs. They enhanced LLM prompts using outputs from PathCrawler and EVA to produce ACSL annotations. Their findings showed that PathCrawler generated context-aware annotations, while EVA contributed to reducing runtime errors. The work \cite{SpecGen2024} represents a novel framework named SpecGen to generate specifications through LLMs using two phases. The first phase is about having prompts in conversational style. The second phase is deployed where correct specifications are not generated. Here, four mutation operators are applied to ensure the correctness of the generated specifications. Two benchmarks i.e. SV-COMP and SpecGen are used. Verifiable specifications are generated successfully for 279 out of 384 programs, making \cite{SpecGen2024} a viable approach.

Related work in Business Process Models (BPMs) is presented in \cite{6823180}. BPMs specify the requirements of process-aware information systems. The automatic generation of natural language requirements which describes these models is presented by the authors. Caglayan et al. \cite{BoraCaglayan2024} deals with the challenges involved in NL2SQL transformation, being widely deployed in Business Intelligence (BI) applications. Caglayan et al. \cite{BoraCaglayan2024} developed a new benchmark focused on typical NL questions in industrial BI scenarios. Authors added question categories in the developed benchmark. Furthermore, two new semantic similarity evaluation metrics are represented in \cite{BoraCaglayan2024}, increasing NL2SQL transformation capabilities. 

\begin{table}[th]
   \caption{Summary of Tools, Frameworks, and Achievements}
   \label{tab:summary1to13}
   \centering
   \begin{tabular}{ r@{} p{0.3cm}  p{6cm} }
     \toprule
     \multicolumn{2}{c}{\textbf{Paper}} & 
                                          \multicolumn{1}{c}{\textbf{Tool / Framework / Technique \& Description}} \\
     \midrule
     \cite{cosler2023nl2specinteractivelytranslatingunstructured}  &  & \textbf{nl2spec:} A framework leveraging LLMs to generate formal specifications from natural language, addressing ambiguity in system requirements with iterative refinement. \\
     \cite{10691792}  &  & \textbf{AssertLLM:} A tool generating assertions for hardware verification from design specifications using three customized LLMs, achieving 89\% correctness. \\
     \cite{li2024specllmexploringgenerationreview}  &  & \textbf{SpecLLM:} Explores the use of LLMs for generating and reviewing VLSI design specifications, aiding chip architects by synthesizing specifications from scratch and from RTL code. \\
     \cite{10207159}  &  & \textbf{Symbolic NLP vs. ChatGPT:} Compares symbolic NLP and ChatGPT performance in generating correct JML output from given natural language preconditions. \\
     \cite{quan2024verificationrefinementnaturallanguage}  &  & \textbf{Explanation-Refiner:} A neuro-symbolic framework integrating LLMs and theorem provers to formalize and validate explanatory sentences, ensuring correctness and providing feedback for improving NLI models. \\
     \cite{Granberry2025a}  &  & \textbf{PathCrawler + EVA:} Combines symbolic analysis and LLMs to generate specifications for C programs, leveraging PathCrawler for context-aware annotations and EVA for reducing runtime errors. \\
     \cite{SpecGen2024}  &  & \textbf{SpecGen:} A framework generating verifiable specifications through conversational prompts and mutation operators, successfully generating 279 correct specifications out of 384 programs. \\
     \cite{6823180}  &  & \textbf{BPM-to-NL Translation Process:} Generated natural language descriptions from business process models for better validation. \\
     \cite{BoraCaglayan2024}  &  & \textbf{NL2SQL Benchmark:} Introduces a new benchmark for NL2SQL transformation in Business Intelligence applications, adding question categories and new semantic similarity evaluation metrics. \\
     \bottomrule
   \end{tabular}
 \end{table}

\section{Conclusions} 

The key contribution of existing work in this area is on bridging the gap between informal natural language descriptions and rigorous formal specifications. In VERIFAI, we aim to improve the techniques that bridge this gap, through  refinement of prompt engineering, the incorporation of chain-of-thought reasoning and the development of hybrid neuro-symbolic approaches.

\section{Acknowledgements}
This work is partly funded by the ADAPT Research Centre for AI-Driven Digital Content Technology, which is funded by Research Ireland through the Research Ireland Centres Programme and is co funded under the European Regional Development Fund (ERDF) through Grant 13/RC/2106 P2. The submission aligns with Digital Content Transformation (DCT) thread of the ADAPT research centre.
\balance
\bibliographystyle{IEEEtran}
\bibliography{adaptConfBib}

\end{document}